# Localized domain-wall excitations in patterned magnetic dots probed by broadband ferromagnetic resonance

F.G. Aliev,[1] A.A. Awad,[1] D. Dielman,[1] A. Lara[1], V. Metlushko,[2] and K.Y. Guslienko[3,4]

[1] *Dpto. Física de la Materia Condensada, CIII, Universidad Autónoma de Madrid, 28049 Madrid, Spain*

[2] *Dept. Electrical & Computer Engineering, University of Illinois at Chicago, Chicago, IL, USA*

[3] *Dpto. Física de Materiales, Universidad del País Vasco, 20018 San Sebastian, Spain*

[4] *IKERBASQUE, The Basque Foundation for Science, 48011 Bilbao, Spain*

We investigate the magnetization dynamics in circular Permalloy dots with spatially separated magnetic vortices interconnected by domain walls (double vortex state). We identify a novel type of quasi one-dimensional (1D) localised spin wave modes confined along domain walls, connecting each of two vortex cores with two edge half-antivortices. Variation of the mode eigenfrequencies with the dot size is in quantitative agreement with the developed model, which considers a dipolar origin of the localized 1D spin waves or so-called Winter´s magnons [J.M. Winter, Phys.Rev. **124**, 452 (1961)]. These spin waves are analogous to the displacement waves of strings, and could be excited in a wide class of patterned magnetic nanostructures possessing domain walls, namely in triangular, square, circular or elliptic magnetic dots.

Vortex forms are found in a variety of physical systems, ranging from superfluids and superconductors to tornadoes. As for the single magnetic vortex (SV) state in confined geometry, its first observation already revealed the coexistence of a SV and domain walls (DW), which connect the SV with the vertices of a triangular nanomagnet [1]. Special interest in the confined magnetic vortices is inspired by the possibility of excitation or switching of the vortex core [2,3] which has been suggested as a potential new road to the creation of nanoscale memory and spin torque vortex oscillators [4]. It is now well established that *coexisting vortex-DW configurations* are formed in the ground states of magnetic rectangles, triangles or ellipses [1,5-7], before the single-domain state appears upon decreasing the particle size [5]. Although excitations of the confined DW or SV states are well understood [8-12], the nature of the *spin waves excited in nanomagnets involving both vortices and domain walls* remains unclear. Circular dots are an example of nanomagnets where eigenmodes in the SV ground state can simply be described analytically [11,13]. Decreasing the external magnetic field from saturation, these nanomagnets are known to have two intermediate metastable states: the double magnetic vortex (DMV) and the so-called S-state, which is formed by a DW only [14]. Recent reports show that circular magnetic dots with reduced thickness, and therefore enhanced pinning, may accommodate the DMV as a long-living metastable state, with two magnetic vortices connected via a DW to edge localized antivortices (AV) [14-17].

Circular dots provide a unique opportunity to investigate the spin wave dynamics of both the vortex-DW (DMV) and SV configurations by reducing or increasing the external magnetic field below or above the vortex nucleation field, $H_n$. In this system, the vortex cores can be pinned by microstructure related defects, intergrain boundaries, etc. [18], thereby trapping the system in the metastable DMV state. Provided the perturbation is small enough, the dot can stay in its metastable state for at least the duration of the measurements. We show here that for low fields the magnetization dynamics of circular Py dots with diameter of 1000 nm and a thickness below 25 nm corresponds to DMV state excitations. The first few lowest frequency spin wave modes present the first example of confined (standing) quasi-1D spin waves localised along domain walls pinned between the vortex and half-antivortex cores.

For the experiments, a vector network analyzer (VNA-FMR) was used to conduct broadband measurements of the magnetization dynamics in the square arrays of Permalloy dots with spacing between dot edges exceeding diameter in order to avoid influence of dipolar interaction. We excited the dots by applying small (0.1 Oe) in-plane driving fields created by a co-planar waveguide (CPW), parallel to the magnetic bias field. This technique ensures small precession amplitudes of the magnetization around the effective field. The data was analyzed on the basis of the transmission model, under the assumption that the dominant CPW mode is TEM mode and by neglecting the effect of reflection [19]. In the experiments we investigated the evolution of the broadband dynamic response of the dot as a function of a slowly changing in-plane magnetic field. We first saturate the sample by applying a positive field that is two times larger than the positive vortex annihilation field ($H_a$). Subsequently we slowly sweep the field downwards, lowering it past the vortex nucleation field ($H_n$) and after changing the field polarity we end at the negative vortex annihilation field. The microwave field drive, parallel to the external bias field, with an amplitude a thousand times smaller than $H_a$, mainly couples to DW where the exchange energy is the highest. This permits precise detection of possible localized excitations in the DW [11]. We investigated circular dots with the thicknesses $L=$ 50, 25, 20 and 15 nm with diameter $2R = 1035$ nm and 20 nm thick dots with $2R=570$ nm. The micromagnetic simulations were carried out using OOMMF code [20] for circular Py dots with a diameter of 1035 nm and thickness of 20 nm, with simulated cell size 5 x 5 x 20 nm$^3$, $\gamma/2\pi = 2.96$ MHz/Oe, the exchange stiffness constant $A = 1.4 \times 10^{-11}$ J/m, and the Gilbert damping constant $\alpha = 0.01$. We first saturate the dot, using an in-plane field of 1000 Oe. After releasing the field in one step to zero, in agreement with [14], the dot relaxes by means of two clearly defined steps that have a slope of energy vs. time of almost zero (not shown). These steps are two metastable magnetic states which we identify as first the S-state and afterwards the DMV state. For our simulations these states remain stable for about 10 ns only. This can be attributed to the fact that we neglected the magnetic anisotropy and all the imperfections that the real samples usually have. To excite magnetization dynamics, a variable driving field (Gaussian field pulse) with the amplitude of 1 Oe and FWHM of 1 ps was applied. We perform local Fourier transforms over all simulation cells and average these spectra to obtain the eigenfrequencies.

Figure 1 compares the static magnetization and dynamic response for dots with thicknesses $L$=50 nm (left) and 20 nm (right). The spin wave resonances in 50 nm thick dots for bias fields below vortex nucleation field $|H_n|$ correspond to the azimuthal modes [19] and are symmetric with respect to the field direction, as expected for the SV state (Fig.1a,c). However, this low field response drastically changes for the Py dots with smaller $L$. Starting from 25 nm the magnetization dynamics becomes asymmetric at low fields (Fig.1d). Besides, we find for a parallel excitation field and for bias fields just below $H_n$, additional peaks with frequencies between 2 and 5 GHz, which are lower than the first azimuthal mode frequency of the SV state, but much higher than the vortex gyrotropic frequency [6]. We claim that these surprising spectra originate from another, intermediate state. Namely, we consider that the Py dots in the array are in a metastable DMV state in the field interval indicated by vertical dashed lines in Fig.1. A reduction of the vortex core size and a decrease in the dot's thickness [21], accompanied by the enhancement of the effective damping, may be an origin of the stabilization of the metastable DMV state for the thinner dots, in agreement with previous measurements [14-17].

In order to investigate the possible metastable states of the dots, we performed micromagnetic simulations. We identified the measured asymmetric eigenmodes (Fig.1d) by comparison of the calculated excitation spectra for different metastable (S or DMV) states with the experimental results. Comparison of the dynamic simulations for the S-type and DMV states with experimental data reveals that the spin wave modes excited between the positive $H_n$ and small negative bias fields, are close to those expected for the DMV state (Fig.2). The robustness of the dynamic response (not shown) further confirms that the detected spin waves are due to the DMV state. Figure 2a shows the three strongest peaks (above the gyrotropic frequency) of the simulated spectra compared to the experiments. Clearly, there is a good match between the experiments and simulations.

In order to get more insight into the spin dynamics in the DMV state we investigate the spatial distribution of the main eigenmodes. Knowing the local distribution of the phases and amplitudes for every cell for selected a eigenfrequency, we can reconstruct the eigenmode profiles [22]. Figure 3 identifies the main eigenmodes excited in the DMV state at zero applied field. The modes are numbered according to the spectra in Fig.2b. The first eigenmode is primarily a gyrotropic mode type excitation of the cores. Modes 2,3,4 and 5 are localized along the domain walls that connect the vortex cores (VC) and edge localized half-antivortex cores. Mode 2 is of an optical type (the different DW´s connecting the V-AV-V-AV rhomb apexes are oscillating out-of-phase) and is localised closer to the dots edge, at the ends of the DWs. Mode 3 shows a node in the oscillation amplitude. The mode 4 (acoustic) reveals no nodes but a strong pinning at the AV positions near the dot edges. Eigenmode 5 is a running wave along V-AV-V-AV rhomb. The higher frequency modes 6 and 7 show strong features localized outside the domain walls. In Supplementary Movies [23] we provide time sequences of the variation of $M_x$ component of magnetization corresponding to the spin wave modes 2-4 for the 1035x20 nm Py dot shown in Fig.3.

We note that in the case of the SV excitations in square dots [7] the VC is a crossing of the four 90-degrees DWs connecting the core and the square corners. The VC position oscillations then correspond to change of the DW shape (DW bulging) and length. The VC mode and DW bulging modes are naturally coupled and have the same oscillation eigenfrequency (230

MHz in [7]). These bulging DW modes are essentially different from the here considered DW flexural modes, having eigenfrequencies about 2 GHz and higher.

There are two branches of the spin excitations in the presence of a DW: ¨free spin waves¨ with excitations resembling those in an infinite uniform ferromagnet, and wall-bounded excitations or Winter´s magnons [24], representing oscillations of the DW shape about the static equilibrium position which can be described as a plane. The DW displacement waves are analogous to the displacements that can propagate along elastic strings (e.g., violin strings) and to the capillary waves on the surface of liquids. The Winter´s magnons were considered for the Bloch type of the DW [24] in an infinite ferromagnet. But in our case of flat soft magnetic dots, the spins lay in-plane due to strong magnetostatic energy and a Néel DW should be considered as a ground state. To understand 1D spin waves confined along the Néel DW that is pinned by the V(AV) cores, we determine spin excitations of the wall applying the approach by Slonczewski [25]. To describe the magnetization dynamics we use the Landau-Lifshitz equation of motion of reduced magnetization $\mathbf{m} = \mathbf{M}/M_s$ ($|\mathbf{M}| = M_s$). We assume that there is a Néel DW connecting the V and AV centers and consider its excitations expressing $\mathbf{m}$ as a sum $\mathbf{m} = \mathbf{m}_0 + \delta\mathbf{m}$ of the DW static magnetization and spin wave contribution. The Cartesian components of $\mathbf{m}$ are defined by the spherical angles $(\Theta, \Phi)$: $m_x + im_y = \sin\Theta \exp(i\Phi)$, $m_z = \cos\Theta$. $(\pi/2, \Phi_0)$ are the spherical angles of the DW ground state $\mathbf{m}_0$. Taking into account that the DW length is about the dot radius $R$ and the dot is thin ($L \ll R$), we apply for the static DW description an infinite film 1D ansatz $\cos\Phi_0(x) = \sec h(x/\delta)$, $\sin\Phi_0(x) = \tanh(x/\delta)$, where $\delta$ is the DW width. Then, introducing substitution $\mathbf{m}(x,y,t) \to \mathbf{m}_0[x - q(y,t)]$ we consider oscillations of the DW shape writing coupled equations of motion for the DW displacement from its rest position $q(y,t)$ and the magnetization deviation angle $\vartheta$ from the dot plane $xOy$ ($\Theta = \pi/2 + \vartheta$). The variables $q$ and $\vartheta$ describe flexural modes of the DW or Winter´s magnons [24]. We introduce a strong pinning near the VC position in the DW ($y=0$) and no pinning or a strong pinning near the AV core ($y=\Delta$, $\Delta \approx R$). This allows using the eigenfunctions $q_n(y) \propto \sin(k_n y)$, with discrete $k_n = (2n-1)\pi/2\Delta$ or $k_n = n\pi/\Delta$, $n=1,2,\ldots$, respectively, and considering the quantized frequencies of the Winter´s magnons $\omega_n = \omega(k_n)$. The magnon frequencies $\omega_n^2/\omega_M^2 = p\int_0^\infty dx x\left(\sqrt{x^2 + k_n^2 \delta^2} - x\right)/\sinh(\pi x) + (4/\pi)\exp(-\Delta/\delta)$, $p = L/\delta$, $k_n L \ll 1$, are in a good agreement with the measured eigenfrequencies (here $\omega_M = 4\pi\gamma M_s$). Applicability of the model is confirmed by considering variation of the eigenfrequencies for modes 2 and 4 with the dot aspect ratio (Fig. 4).

In conclusion, we identify quasi one-dimensional (1D) localised spin wave modes confined along the domain walls, connecting vortex cores with half-antivortices in circular magnetic dots. In general, *similar spin waves would be excited by spin torque or microwave magnetic filed in thin circular, triangular or rectangular patterned magnetic nanostructures possessing magnetic domain walls.* Besides, our findings would lead to creation of principally new magnetic devices. For example, the mode 3 with oscillating magnetic charges at the edges (Fig. 3) would be used to develop microwave nano-emitters. Finally, our results introduce a scheme to investigate the high-frequency magnetization dynamics of symmetry breaking of metastable states in highly symmetric patterned magnetic elements.

**References**

[1] A. Tonomura, et al., *Phys. Rev. Lett.,* **44**, 1430 (1980).

[2] K. Yamada, et al. *Nature Mat.* **6**, 269 (2007).

[3] B. Van Waeyenberge et al., *Nature* **444**, 461 (2006).

[4] A. Dussaux, et al. *Nature Comm.* **1**, 1 (2010).

[5] M. Hehn, et al., *Science,* **272** 1782 (1996).

[6] K. S. Buchanan, et al., *Nature Phys.* **1**, 172 (2005).

[7] J. Raabe et al., *Phys Rev. Lett*. **94**, 217204 (2005).

[8] E. Saitoh, et. al, *Nature*, **432**, 203 (2004).

[9] L. Thomas, et al., *Nature* **443**, 197 (2006).

[10] D. Bedau, et al., *Phys. Rev., Lett.* **99**, 146601 (2007).

[11] K. Guslienko, et al.*, J. Appl. Phys.* **91**, 8037 (2002).

[12] M. Buess, et al., Phys. Rev. Lett. **93**, 077207 (2004).

[13] A.A. Awad, et al.*, Appl. Phys. Lett.* **96**, 012503 (2010).

[14] C. Vaz, et al., *Phys. Rev.* B**72**, 224426 (2005).

[15] T. Pokhil, D. Song, J. Nowak, *J. Appl. Phys* **87**, 6319 (2000).

[16] M. Rahm, et al*. Appl. Phys. Lett.* **82**, 4110 (2003).

[17] I. Prejbeanu, et al. *J. Appl. Phys.* **91**, 7343 (2002).

[18] R. Compton, P. Crowell, *Phys. Rev. Lett.* **97**, 137202 (2006).

[19] F. G. Aliev, et al., *Phys. Rev. B* **79**, 174433 (2009).

[20] Object oriented micromagnetic framework. math.nist.gov/oommf/software.html.

[21] A. Hubert, and R. Schäfer, *Magnetic Domains* (Springer, 1998).

[22] R. McMichael, M. Stiles, *J. Appl. Phys.* **97**, 10J901 (2005).

[23] See EPAPS Documents No. [number will be inserted by publisher] for [Videos 1-3. Time dependent variation of the $M_x$ component of magnetization corresponding to modes 2-4 (1035x20nm Py dot)].

[24] J. M. Winter, Phys. Rev. **124**, 452-459 (1961).

[25] J. C. Slonczewski, *J. Magn. Magn. Mat.*, **31-34**, 663-664 (1983).

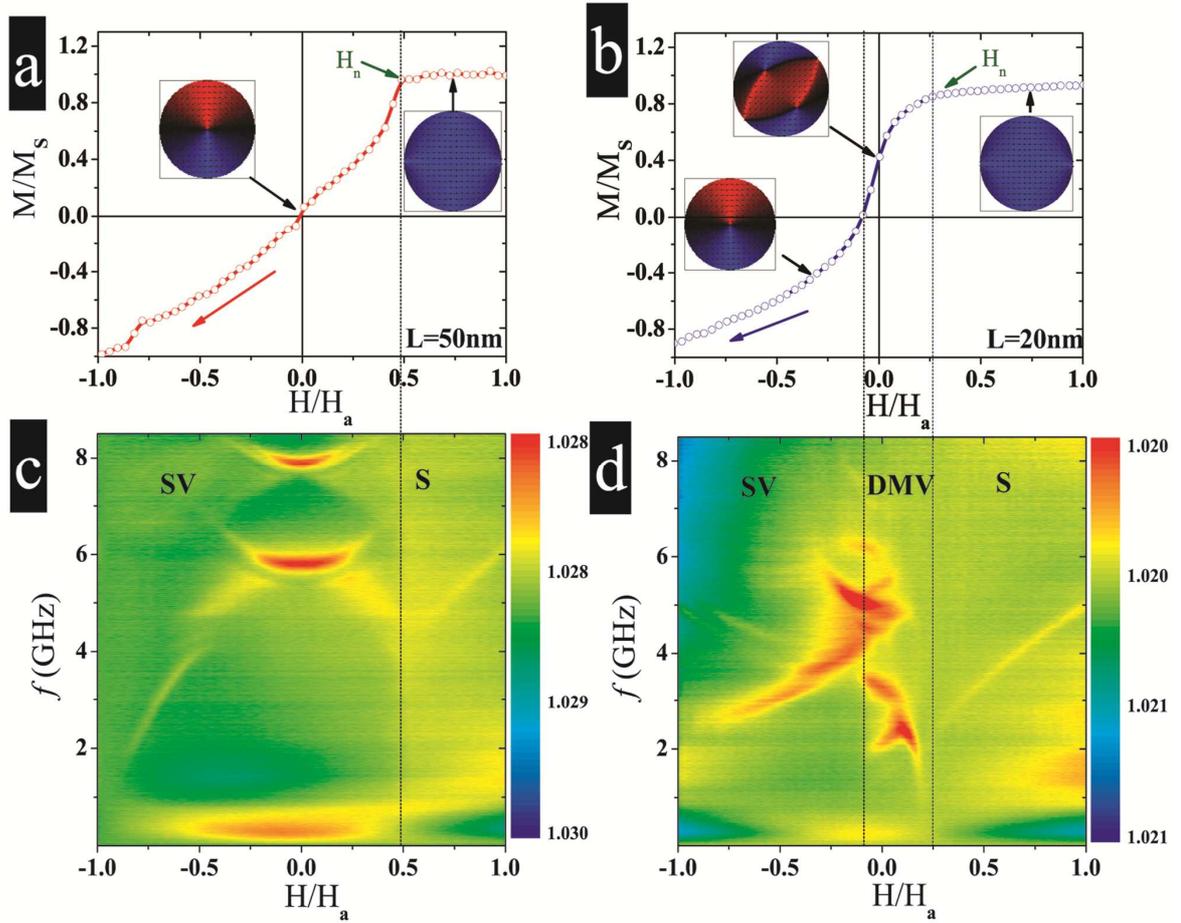

Figure 1: (a,b) show the static magnetization of the Py dots with 1000 nm diameter and thicknesses of 50 and 20 nm respectively measured by sweeping field from $H_a$ to $-H_a$. Parts (c,d) show intensity plots of the measured spin-wave spectra for the Py dot arrays with the bias field swept down from saturation and the driving field applied parallel to the bias field. Magnetic field is normalized by the vortex annihilation field $H_a$. The vortex nucleation fields are marked as $H_n$, inserts and vertical lines remark saturated (SAT), DMV and SV magnetic states.

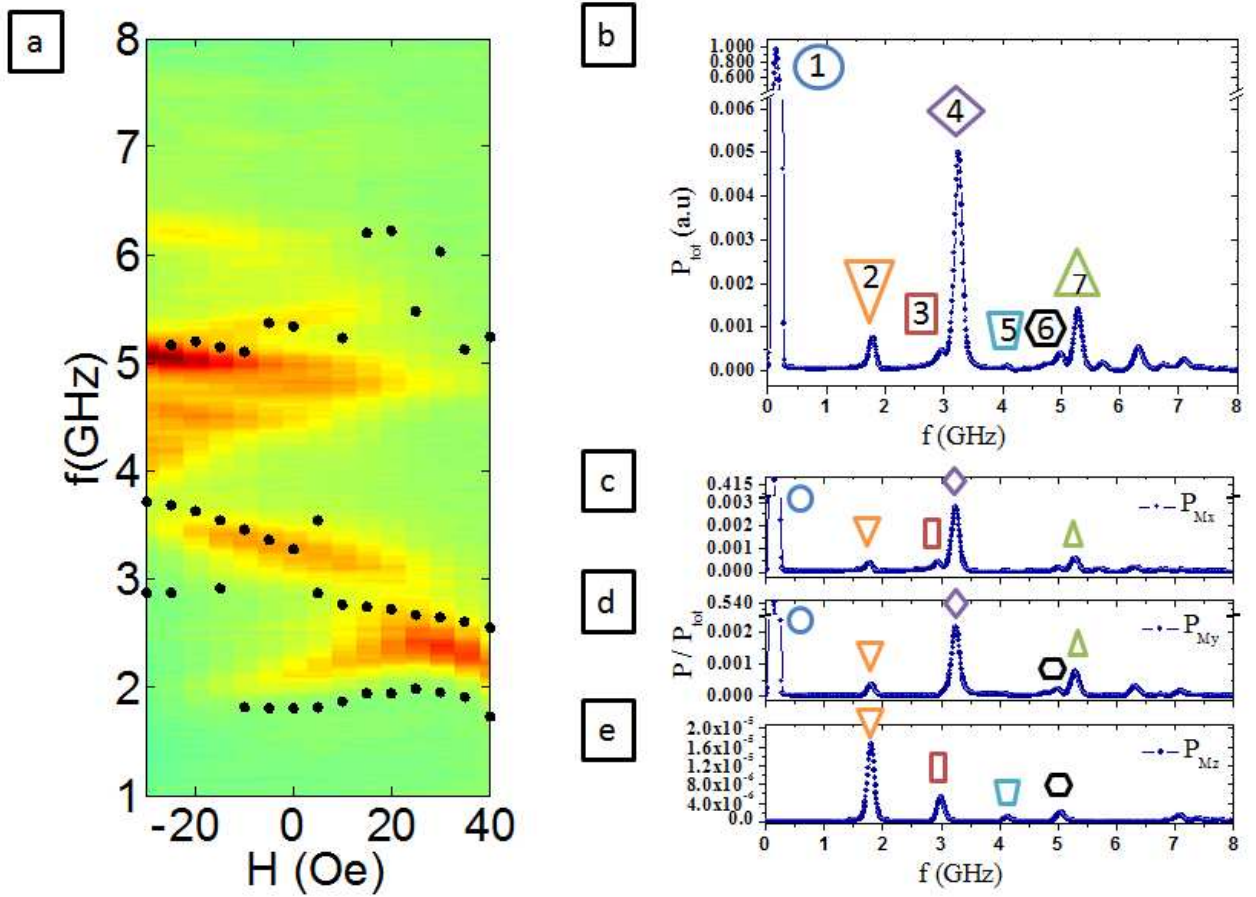

Figure 2: Left: Comparison of the simulations (closed circles) with the experiments (background, see marked part of Fig. 1d). The simulated frequencies are obtained by taking the three strongest peaks at each applied field. Right: (b) Simulated spectrum at zero field for all the magnetization components $M_x$, $M_y$ and $M_z$ combined and the components separately, respectively (c), (d) and (e). In (b) the intensity peaks are marked by integers, corresponding to the first 7 excited eigenmodes (see Fig. 4 for the spatial distributions of these modes). The dot sizes for both the simulations and experiment are $2R=1035$ nm and $L=20$ nm.

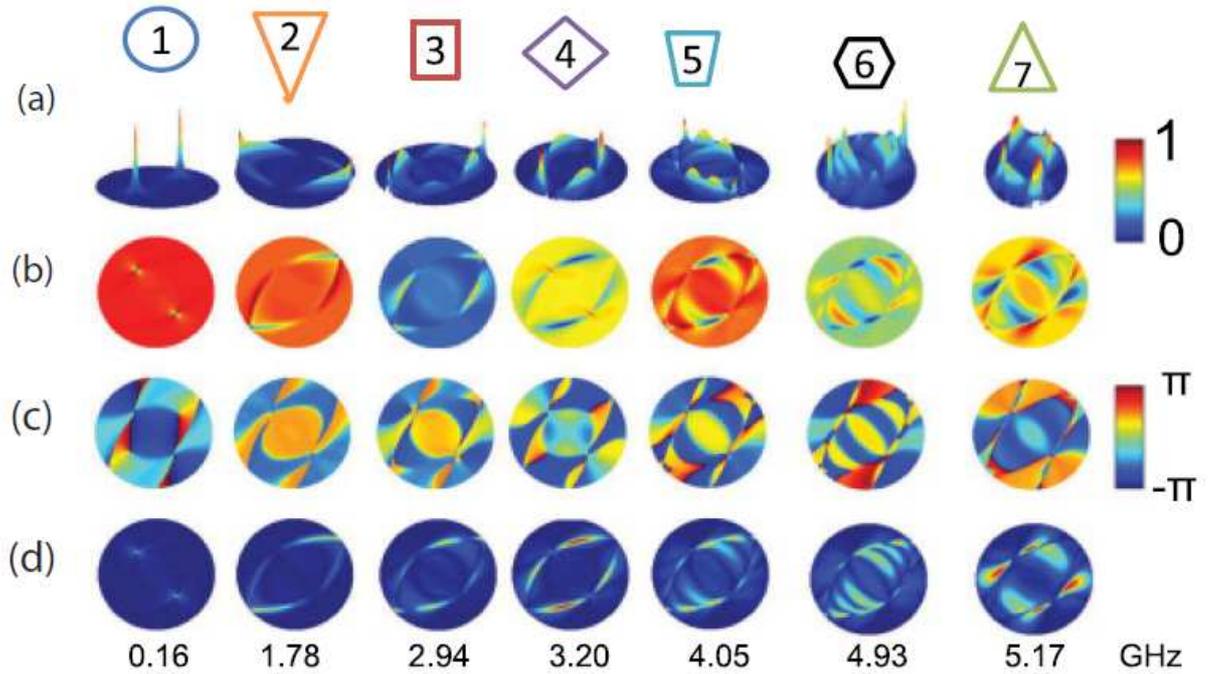

Figure 3: The spatial distributions of the first 7 eigenmodes at zero bias field corresponding to the frequency spectra of Fig. 3. The snapshots for the relative magnetization $\Delta M_x$ component are as follows: (a) 3D-visualization of modulus (a.u.), (b) the real part (a.u.), (c) the phase and (d) 2D-visualization of the amplitude (a.u.). The dot sizes are $2R=1035$ nm and $L=20$ nm. Supplementary Movies [23] show online 3D animations of eigenmodes 2,3 and 4.

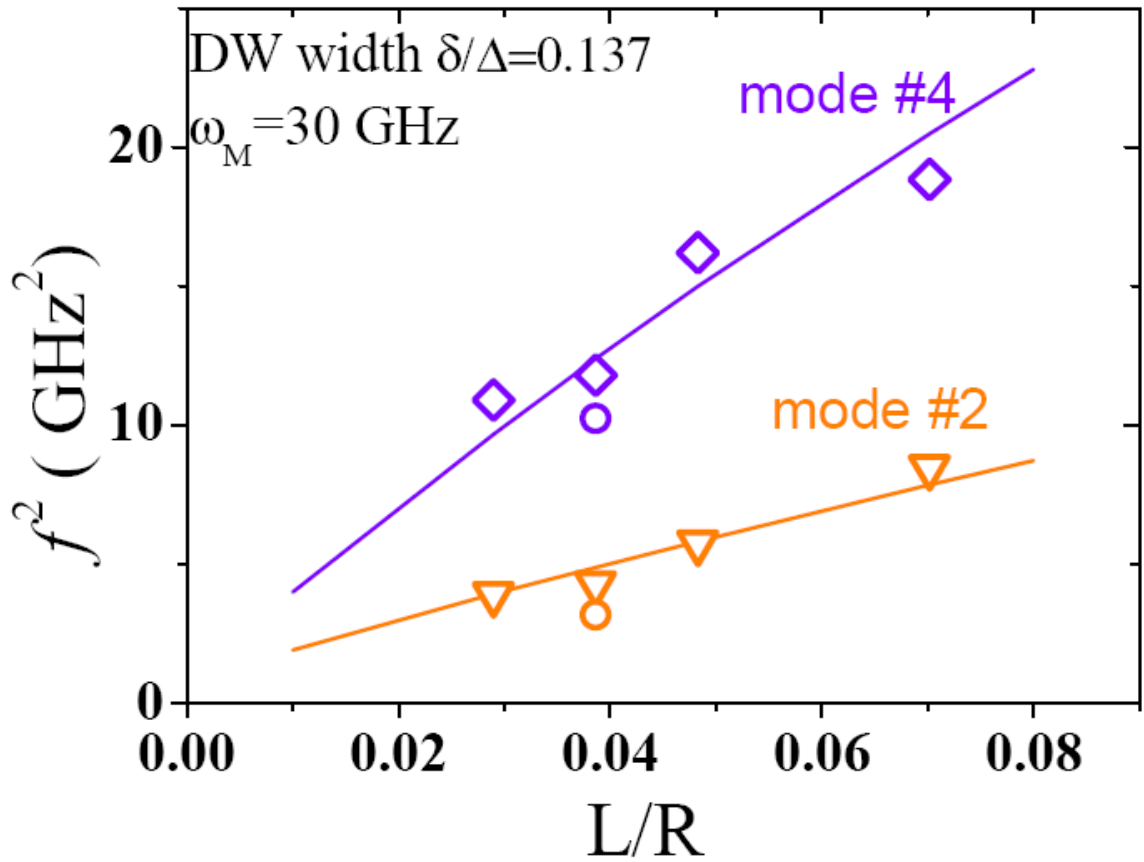

Figure 4 Comparison of the experimental (closed dots), analytical (lines) and simulated (open dots) eigenfrequencies for the most intensive modes #2 and #4 as function of the dot aspect ratio (thickness over radius).